\documentclass{article}
\usepackage{spconf,amsmath,graphicx,hyperref}
\usepackage{url}
\usepackage{graphicx}
\usepackage{cite}
\usepackage{epsf}
\usepackage{float}
\usepackage{times}
\usepackage{color}
\usepackage{cuted}
\usepackage{stfloats}
\usepackage{tabularx}
\usepackage{graphicx}
\usepackage{arydshln}
\usepackage{amsmath}
\usepackage{amssymb}
\usepackage{wasysym}
\usepackage{multirow}
\usepackage{comment}
\usepackage{mathtools}
\usepackage{subeqnarray}
\usepackage{hyperref}
\usepackage{bm}

\usepackage{algorithmic}
\usepackage[ruled]{algorithm}

\makeatletter
\g@addto@macro\normalsize{%
  \setlength\abovedisplayskip{3pt}%
  \setlength\belowdisplayskip{3pt}%
  \setlength\abovedisplayshortskip{2pt}%
  \setlength\belowdisplayshortskip{2pt}%
}
\makeatother

\definecolor{mycolor}{RGB}{0, 0, 255}

\graphicspath{{figures/}}
\ninept
\usepackage[font=small,skip=8pt]{caption} 
\makeatletter

\renewcommand\section{%
  \@startsection{section}{1}{0pt}%
    {-6pt} 
    {6pt}   
    {\normalfont\large\bfseries}%
}

\renewcommand\subsection{%
  \@startsection{subsection}{2}{0pt}%
    {-6pt} 
    {6pt}  
    {\normalfont\normalsize\bfseries}%
}
\makeatother

\title{\bf Joint Optimization of Microphone Array Geometry, Sensor Directivity Pattern, and Beamforming Parameters for Linear Superarrays}
\name{\centerline{\it 
Yuanhang Qian$^{1}$, Xueqin Luo$^{2}$, Jilu Jin$^{2}$,  Gongping Huang$^{1}$, Jingdong Chen$^{2}$, and Jacob Benesty$^{3}$}
}

\address{\centerline{$^{1}$Electronic Information School of Wuhan University, 430072, Wuhan, China.} \\
\centerline{$^{2}$CIAIC, Northwestern Polytechnical University, Xi'an, Shaanxi 710072, China.} \\
\centerline{$^{3}$INRS-EMT, University of Quebec, Montreal, QC H5A 1K6, Canada.}
}

\begin{document}
\maketitle

\begin{abstract}
Linear superarrays (LSAs) have been proposed to address the limited steering capability of conventional linear differential microphone arrays (LDMAs) by integrating omnidirectional and directional microphones, enabling more flexible beamformer designs. However, existing approaches remain limited because array geometry and element directivity, both critical to beamforming performance, are not jointly optimized. This paper presents a generalized LSA optimization framework that simultaneously optimizes array geometry, element directivity, and the beamforming filter to minimize the approximation error between the designed beampattern and an ideal directivity pattern (IDP) over the full frequency band and all steering directions within the region of interest. The beamformer is derived by approximating the IDP using a Jacobi–Anger series expansion, while the array geometry and element directivity are optimized via a genetic algorithm. Simulation results show that the proposed optimized array achieves lower approximation error than conventional LSAs across the target frequency band and steering range. Additionally, its directivity factor and white noise gain demonstrate more stable and improved performance across frequencies and steering angles.
\end{abstract}


\section{Introduction}
\label{Sect-Intro}

Microphone arrays combined with beamforming techniques have been extensively used in a wide range of applications~\cite{brandstein2001microphone, elko2008microphone, benesty2017fundamentals}. In designing microphone arrays, two key factors significantly impact beamforming performance: the array geometry~\cite{itzhak2024joint, moisseev2024array} and the selection of sensor elements~\cite{d2006vector, smith2007steering, hawkes1998acoustic}. Common array geometries include linear arrays~\cite{benesty2008microphone, benesty2012study, chen2014design, luo2021constrained}, planar arrays~\cite{benesty2015design, huang2017design, zou2009circular}, and volumetric arrays~\cite{yan2010optimal, rafaely2015fundamentals}. Among these, linear arrays are particularly favored in practice because of their simplicity and ease of integration into devices.

When combined with differential beamforming, microphone arrays can achieve high directivity and nearly frequency-invariant spatial responses, making them well-suited for broadband speech acquisition. Such configurations are commonly referred to as linear differential microphone arrays (LDMAs)~\cite{elko1997steerable, huang2020robust, borra2020efficient}. However, conventional first-order LDMAs are limited to forming a fixed mainlobe in the endfire direction~\cite{bernardini2017wave, huang2020simple, borra2019uniform}. While higher-order LDMAs~\cite{byun2018continuously} can partially steer the beam, their performance rapidly degrades as the steering angle deviates from the endfire~\cite{jin2021steering}. To address this limitation, our recent work~\cite{luo2023design} introduced linear superarrays (LSAs), which combine omnidirectional and bidirectional microphones~\cite{olson1946gradient}, enabling robust beamforming across a wider range of steering directions. However, that design was restricted to arrays composed solely of these two microphone types. More recently, this approach was generalized in~\cite{luo2024design}, where the LSA beampattern is decomposed into two parts: one from an omnidirectional subarray and another from a directional subarray that can include microphones beyond just bidirectional types. Although this generalization allows the use of microphones with different directivity patterns, it does not account for array geometry optimization, which may limit the performance.

To address the limitations of existing designs, this paper proposes a generalized LSA optimization framework that simultaneously optimizes array geometry, sensor directivity patterns, and beamforming filters. Unlike prior work, we treat the microphone directivity patterns as design variables. While such configurable sensors may not yet be commercially available, the results offer valuable insights for future sensor development.

Within this framework, the array geometry, sensor directivity patterns, and beamformer coefficients are jointly optimized to minimize the approximation error between the synthesized beampattern and a desired ideal beampattern (IDP) over the full frequency band and a defined range of steering directions~\cite{chen2007optimal}. The beamformer is designed using a Jacobi-Anger series expansion~\cite{benesty2015design}, while the geometry and directivity patterns are optimized via a genetic algorithm~\cite{de1988learning, deep2009real}. Simulation results show that the proposed method outperforms existing LSA designs, achieving lower approximation error and offering more stable and improved performance in terms of directivity factor (DF) and white noise gain (WNG) across the target frequency band and steering range.

\section{Signal Model and Performance Metrics}

We consider an LSA comprising $M$ microphones positioned along the $x$-axis. Each microphone is assumed to have a frequency-invariant directivity pattern, with the pattern of the $m$th microphone being
\begin{equation}
    \mathcal{A}_m(\theta) = a_{m} + \bigl(1 - a_{m}\bigr)\sin\theta ,
\end{equation}
where $\theta$ denotes the azimuth angle measured counterclockwise from the positive $x$-axis, and $a_{m}$ is a parameter that characterizes the directivity of the $m$th microphone. Specifically, $a_{m}=1$ corresponds to an omnidirectional pattern, while $a_{m} \neq 0$ indicates a directional microphone with its main lobe oriented toward the positive $y$-axis.

Given the described LSA configuration, the $m$th element of the array manifold vector $\mathbf{d} \left( \mathbf{x}, \theta, \omega \right)$, is expressed as~\cite{van2002optimum, benesty2008microphone}  
\begin{equation}
\label{sv}
    \left[ \mathbf{d} \left(\mathbf{x}, \theta, \omega \right) \right]_m 
    =  \mathcal{A}_m \left( \theta \right)  
    e^{\jmath \varpi x_m \cos \theta }, 
    \quad m = 1, 2, \ldots, M ,
\end{equation}
where $\varpi = \omega / c$, $\omega = 2 \pi f$, $f$ denotes the frequency, and $c$ is the speed of sound. The vector $\mathbf{x} = \left[ x_1~~ x_2~~ \cdots~~ x_M \right]^T$ contains the positions of all microphones along the $x$-axis, where each $x_m \in \left[0, L \right]$ represents the coordinate of the $m$-th microphone and $[0, L]$ defines the array's admissible aperture.

Consider a desired signal arriving from direction $\theta_\mathrm{s}$. For a beamformer designed based on the LSA, the following distortionless response constraint must be satisfied: 
\begin{equation}
\label{const-oo}
    \mathbf{h}^H \left( \omega \right) 
    \mathbf{d} \left( \mathbf{x}, \theta_\mathrm{s}, \omega \right) = 1 ,
\end{equation}
where $\mathbf{h} \left( \omega \right) \in \mathbb{C}^M$ denotes the beamforming filter and the superscript ${(\cdot)}^H$ stands for the conjugate-transpose operator.

Three commonly used metrics are outlined below to assess the performance of the beamformer.
\begin{itemize}
\item
Beampattern\cite{benesty2008microphone}, which is defined in our context as
\begin{align}
\label{beam}
\mathcal{B} \left[ \mathbf{h} \left( \omega \right),  \mathbf{x},  \theta \right] 
 =  \mathbf{h}^H \left( \omega \right) \mathbf{d} \left( \mathbf{x} , \theta, \omega \right).
 \vspace{-2pt}
\end{align}

\item
WNG \cite{brandstein2001microphone}, which is expressed as
\begin{align}
\label{gain-SNR-WNG}
\mathcal{W} \left[\mathbf{h} \left( \omega \right), \mathbf{x} \right]
= \frac{\left| \mathbf{h}^H \left( \omega \right) \mathbf{d} \left(\mathbf{x}, \theta_{\mathrm{s}}, \omega \right) \right|^2}{\mathbf{h}^H \left(  \omega \right)  \mathbf{h}\left( \omega \right) }.
\end{align}
\item
According to the definition in~\cite{elko2004differential, elko2008microphone}, the DF of the LSA in two-dimensional (2D) space can be written as
\begin{align}
\label{gain-SNR-DF}
\mathcal{D} \left[ \mathbf{h}\left(\omega \right), \mathbf{x} \right]
&=
 \frac{\left| \mathbf{h}^H \left( \omega \right) \mathbf{d} \left( \mathbf{x}, \theta_{\mathrm{s}}, \omega \right) \right|^2} {\mathbf{h}^H \left( \omega \right) \mathbf{\Gamma} \left( \mathbf{x}, \omega \right) \ \mathbf{h}\left( \omega \right)},
\end{align}
where the elements of $\mathbf{\Gamma} \left( \mathbf{x}, \omega \right)$ can be derived following the methods outlined in \cite{luo2023design, luo2024design} as
\begin{align}
\label{Gamma-ij}
\left[ \mathbf{\Gamma} \left( \mathbf{x}, \omega \right)  \right]_{ij}
& = \left(a_{0,i} a_{0,j} + \frac{a_{1,i} a_{1,j}}{2} \right) J_0 \left( \varpi \Delta x_{ij} \right)  \nonumber\\
& \quad + \frac{a_{1,i} a_{1,j}}{2} J_2 \left( \varpi \Delta x_{ij} \right), 
\end{align}
with $\Delta x_{ij}  = \left| x_i - x_j \right|, \  i, j  = 1, 2, \ldots, M $, and $J_n\left(\cdot\right)$ denoting the $n$th-order Bessel function of the first kind satisfying $J_n\left(\cdot \right) = \left(-1\right)^n J_{-n}\left(\cdot\right)$.
\end{itemize}

\section{Problem Formulation}

The primary goal in designing an LSA is to implement a two-dimensional, steerable differential beamformer constrained by a linear array geometry, utilizing microphones of various types. To achieve this, an $N$th-order beamformer is constructed so that its beampattern closely approximates the IDP, expressed as~\cite{huang2017design, benesty2012study}
\begin{align}
\label{ideal-beam}
\mathcal{B}_{N} \left( \theta_\mathrm{s}, \theta \right)= \sum^{N}_{n = 0} \alpha_{N, n} \cos \left( n \theta - n \theta_\mathrm{s} \right),
\end{align}
where the coefficients $\alpha_{N, n}$ sum to one. 

The approximation error between the beampattern and the IDP, relative to the steering direction $\theta_\mathrm{s}$ is defined as
\begin{align}
\label{approximating error}
& \epsilon_{N} \left[ \mathbf{h} \left( \omega \right),  \mathbf{x}, \theta_\mathrm{s} \right] \nonumber\\
&\quad= \frac{1}{2\pi}\int^{2\pi}_{0}\left| \mathcal{B} \left[ \mathbf{h} \left( \omega \right),  \mathbf{x},  \theta \right] -
\mathcal{B}_{N}\left(\theta_\mathrm{s}, \theta \right) \right|^2 d \theta.
\end{align}

Assuming that the steering direction of interest $\theta_\mathrm{s}$ lies within the range $\left[\theta_1, \theta_2 \right]$, the overall approximation error across all steering directions and the entire broadband frequency range  $[\omega_\mathrm{L}, \omega_\mathrm{H}]$ is defined as
\begin{align}
\label{overall_error}
\overline{\epsilon}_N \left[ \mathbf{h} \left(  \omega \right) , \mathbf{x} \right] = \int^{\theta_2}_{\theta_1} \int^{\omega_\mathrm{H}}_{\omega_\mathrm{L}} \epsilon_{N} \left[ \mathbf{h} \left( \omega \right),  \mathbf{x}, \theta_\mathrm{s} \right]  \ d \omega \ d \theta_\mathrm{s}.
\end{align}

The goal of our task is then to determine the optimal array geometry  $\mathbf{x}$ and the optimal sensor directivity parameters $a_{m}$'s for each microphone in order to minimize the overall approximation error $\overline{\epsilon}_N \bigl[ \mathbf{h}(\theta_\mathrm{s}, \omega), \mathbf{x} \bigr]$. Formally, the optimization problem can be stated as
\begin{align}
\label{opt}
& \left(\hat{\mathbf{x}},\hat{\mathbf{a}}\right)=\min_{\mathbf{x}, \mathbf{a}, \mathbf{h} \left( \omega \right)} \overline{\epsilon}_N \left[ \mathbf{h} \left(  \omega \right) , \mathbf{x} \right] \nonumber\\
&~~~~~~~~~~~~~~~~ \text { s. t. }\left\{\begin{array}{l}
\left|x_i-x_j \right| \geq d_{c}   \\
x_{i} \in \left[0, L \right] \\
1 \leq i, j \leq M,  \ i \neq j
\end{array}\right. ,
\end{align}
where the vector $\mathbf{a} = \left[ a_{1}~~ a_{2}~~ \cdots~~ a_{M} \right]^T$ specifies the directivity of each array element, and  $\hat{\mathbf{x}}$ and $\hat{\mathbf{a}}$ denote the optimized array geometry and microphone directivity patterns, respectively.

\section{Generalized LSA Optimization Framework}
\subsection{Beamforming Filter Optimization }
Given an array geometry parameter vector $\mathbf{x}$ and the array element directivity parameter $\mathbf{a}$, we propose a method in this section to determine the optimal $\mathbf{h} \left( \omega \right)$ that minimizes $\epsilon_{N} \left[ \mathbf{h} \left( \omega \right),  \mathbf{x}, \theta_\mathrm{s} \right]$. Unlike previous work~\cite{luo2023design,luo2024design}, the proposed beamformer design is more general and can be applied to linear arrays composed of arbitrary combinations of microphones with different directivity patterns.

Following the approach \cite{luo2023design, luo2024design}, we rewrite (\ref{ideal-beam}) as
\begin{align}
\label{ideal-exp}
\mathcal{B}_{N} \left( \theta_\mathrm{s}, \theta \right)
& = \sum^{N}_{n = -N} \eta_{n} \left( \theta_\mathrm{s} \right) e^{\jmath n \theta}\nonumber\\
& \quad +  \sin \theta \sum^{N-1}_{n' = -N+1} \widetilde{\eta}_{ n'} \left( \theta_\mathrm{s} \right)  e^{\jmath n' \theta},
\end{align}
where 
\begin{align}
\displaystyle \eta_{n} \left( \theta_\mathrm{s} \right)&=
\left\{ \begin{array}{ll}
\alpha_{N,\left|n\right|}\cos\left(n\theta_\mathrm{s} \right)/2 ,  & n\neq 0  \\
\alpha_{N,0},  & n=0
\end{array} \right.,\\
\widetilde{\eta}_{n'} \left( \theta_\mathrm{s} \right)& = \sum^{\lfloor \frac{N-1-n'}{2}\rfloor}_{k = \mathrm{max}\{-n',0\}}\gamma_{n'+2k} \left( \theta_\mathrm{s} \right) \binom{k}{n'+2k}2^{-n'-2k},
\end{align}
the operator $\lfloor \cdot \rfloor$ denotes the floor function, which returns the greatest integer less than or equal to its argument, and
\begin{align}
\gamma_{i} \left( \theta_\mathrm{s} \right) & = \sum^{\lfloor \frac{N-1-i}{2}\rfloor}_{k = 0} (-1)^k  \alpha_{N,i+2k+1}\nonumber\\
& \quad \times \sin\left[\left(i+2k+1 \right) \theta_\mathrm{s} \right]2^{i}\binom{k}{i+k}, 
\end{align}
with $i = 0, 1, \ldots, N-1$.

By leveraging the Jacobi-Anger expansion on (\ref{beam}), which is a method recognized for its optimal least-squares approximation of exponential functions in beamforming applications \cite{abramowitz1970handbook, huang2017design}, we derive the following expression:
 \begin{align}
\label{beam-approx}
\mathcal{B} \left[ \mathbf{h} \left( \omega \right),  \mathbf{x},  \theta \right] 
&\approx \sum^{\mathcal{N}}_{n = -\mathcal{N}} e^{\jmath n \theta} \mathbf{h}^H \left( \omega \right) \mathbf{\Lambda}_0 \bm{\beta}_{n}  \left( \omega \right) \\
& + \sin \theta \sum^{\mathcal{N}}_{n =-\mathcal{N}} e^{\jmath n \theta} \mathbf{h}^H \left( \omega \right) \mathbf{\Lambda}_1  \bm{\beta}_{n}  \left( \omega \right) , \nonumber
\end{align}
where $\mathbf{\Lambda}_0 = \mathrm{diag} \left\{ \mathbf{a} \right\}$, $\mathbf{\Lambda}_1 = \mathbf{I} - \mathbf{\Lambda}_0 $, with $\mathbf{I}$ being the identity matrix of size $M \times M$,
$\bm{\beta}_{n}\left( \omega \right) = \left[ \begin{array}{ccc}
\beta_n \left(\varpi x_1 \right)& \cdots & \beta_n\left(\varpi x_M \right)
\end{array} \right]^T$, 
with $\beta_n \left(\cdot \right) = \jmath^n J_n \left(\cdot \right)$, and $\mathcal{N}$ is the truncation order.

By comparing the beampattern in (\ref{beam-approx}) to the IDP in (\ref{ideal-exp}), and using the symmetry property $\beta_{-n} \left( \cdot \right) = \beta_n \left( \cdot \right), \ n = 0, 1, \ldots, \mathcal{N}$, we arrive at the following linear system of equations:
\begin{align}
\label{linear-sys}
\underline{\mathbf{\Theta}}_{\mathcal{N}}\left( \omega \right)  \mathbf{h} \left( \omega \right) 
& = \underline{\bm{\eta}}_{\mathcal{N}, N} \left( \theta_\mathrm{s} \right) ,
\end{align}
where
\begin{align}
\label{Omega-mat}
\underline{\mathbf{\Theta}}_{\mathcal{N}} \left( \omega \right)
& = \left[ \begin{array}{c}
\mathbf{\Theta}_{\mathcal{N}} \left( \omega \right) \mathbf{\Lambda}_0   \\
\mathbf{\Theta}_{\mathcal{N}} \left( \omega \right) \mathbf{\Lambda}_1  
\end{array} \right],
\end{align}
with
\begin{align}
\label{Psi-mat-d}
\mathbf{\Theta}_{\mathcal{N}} \left( \omega \right) 
& = \left[ \begin{array}{cccc}
\bm{\beta}_{0} \left( \omega \right)  & \bm{\beta}_{1}\left( \omega \right)  & \cdots &\bm{\beta}_{\mathcal{N}}\left( \omega \right) 
\end{array} \right]^H,
\end{align}
and
\begin{align}
\label{zeta-vect}
\underline{\bm{\eta}}_{\mathcal{N}, N} \left(\theta_\mathrm{s} \right) 
& = \left[ \begin{array}{cc}
\bm{\eta}^T_{\mathcal{N}, N} \left(\theta_\mathrm{s} \right)   & \widetilde{\bm{\eta}}^T_{\mathcal{N}, N-1} \left(\theta_\mathrm{s} \right) 
\end{array} \right]^T
\end{align}
is a vector of length $2\mathcal{N}+2$, with
\begin{align}
\label{eta-vect-o}
\bm{\eta}_{\mathcal{N}, N} \left(\theta_\mathrm{s} \right) 
&= \left[ \begin{array}{cccccc}
\eta_{0} \left(\theta_\mathrm{s} \right) & \cdots & \eta_{N} \left(\theta_\mathrm{s} \right) & 0 & \ldots & 0
\end{array} \right]^T,\\
\label{zeta-vect-d}
\widetilde{\bm{\eta}}_{\mathcal{N}, N-1} \left(\theta_\mathrm{s} \right) 
&= \left[ \begin{array}{ccccccc}
\widetilde{\eta}_{0} \left(\theta_\mathrm{s} \right) & \cdots & \widetilde{\eta}_{N-1} \left(\theta_\mathrm{s} \right) 
& 0 & \ldots & 0 \end{array} \right]^T.
\end{align}

Given that $M \geq  2\mathcal{N}+2$, the solution to (\ref{linear-sys}) can be derived as
\begin{align}
\label{filter-mn-d}
\mathbf{h}_\mathrm{opt} \left( \omega \right)  = \underline{\mathbf{\Theta}}^H_{\mathcal{N}} \left( \omega \right)   \left[ \underline{\mathbf{\Theta}}_{\mathcal{N}} \left( \omega \right)  \underline{\mathbf{\Theta}}^H_{\mathcal{N}} \left( \omega \right)  \right]^{-1} \underline{\bm{\eta}}_{\mathcal{N}, N} \left(\theta_\mathrm{s} \right) .
\end{align}
Substituting (\ref{filter-mn-d}) into (\ref{opt}), the original optimization problem can be simplified as
\begin{align}
\label{opt-2}
& \left(\hat{\mathbf{x}},\hat{\mathbf{a}}\right)=\min_{\mathbf{x}, \mathbf{a}} \overline{\epsilon}_N \left[ \mathbf{h}_\mathrm{opt} \left(  \omega \right) , \mathbf{x} \right] ~~\text { s. t. }\left\{\begin{array}{l}
\left|x_i-x_j \right| \geq d_{c}   \\
x_{i} \in \left[0, L \right] \\
1 \leq i, j \leq M,  \ i \neq j
\end{array}\right. .
\end{align}

\subsection{Genetic Algorithm for Array Optimization}
We now consider using a genetic algorithm (GA) as an optimizer to solve (\ref{filter-mn-d})~\cite{chen2021planar, yu2013optimal, iseki2015optimization}. Specifically, the GA begins by randomly generating a population of candidate array configurations. Each individual in the population represents a microphone position vector $\mathbf{x}$ and directivity parameter vector $\mathbf{a}$. The fitness of each candidate is evaluated based on its ability to minimize the overall approximation error $\overline{\epsilon}_N \left[ \mathbf{h}_\mathrm{opt}(  \omega) , \mathbf{x} \right]$ as defined in (\ref{opt-2}). Through iterative operations of selection, crossover, and mutation, the GA produces new configurations and effectively explores the solution space, helping to avoid local optima. The overall structure of the algorithm is summarized in Algorithm~\ref{alg:ga_optimization}.

\begin{algorithm}[t!]
\caption{Genetic Algorithm for Joint Optimization of Array Geometry and Element Directivity Parameters}
\label{alg:ga_optimization}
\begin{algorithmic}[1]
\STATE \textbf{Input:} number of microphones $M$, minimum spacing $d_c$, aperture limit $L$, frequency range $[\omega_\mathrm{L}, \omega_\mathrm{H}]$, beamformer order $N$, parameters of IDP $\underline{\bm{\eta}}_{\mathcal{N},N}(\theta_\mathrm{s})$ for $\theta_\mathrm{s} \in [\theta_1,\theta_2]$, population size $N_p$, and number of generations $N_i$.
\STATE \textbf{Output:} optimal array configuration parameters $(\hat{\mathbf{x}},\hat{\mathbf{a}})$.
\STATE \textbf{Initialization:} initial population $\mathcal{P}_0$ of $N_p$ individuals, each represented by $(\mathbf{x},\mathbf{a})$ with
\begin{equation*}
x_m \in [0,L], \quad a_{m} \in [0,1], \quad |x_i-x_j|\ge d_c,
\end{equation*}
where $m=1,\dots,M$.
\FOR{$k=1$ to $N_i$}
    \FOR{each individual $(\mathbf{x},\mathbf{a}) \in \mathcal{P}_{k-1}$}
        \STATE Compute the beamforming filter $\mathbf{h}_{\text{opt}}(\omega)$ via (\ref{filter-mn-d}).
        \STATE Evaluate the approximation error $\epsilon_N[\mathbf{h}_{\text{opt}}(\omega),\mathbf{x},\theta_\mathrm{s}]$ (discrete form of (\ref{approximating error})).
        \STATE Compute fitness as the overall approximation error $\overline{\epsilon}_N[\mathbf{h}_{\text{opt}}(\omega),\mathbf{x}]$ (discrete form of (\ref{overall_error})).
    \ENDFOR
    \STATE Rank the individuals based on their fitness scores and identify the top-performing candidate.
    \STATE Apply selection, crossover, and mutation operators to generate the next population $\mathcal{P}_k$.
\ENDFOR
\STATE \textbf{Return:} the best solution $(\hat{\mathbf{x}},\hat{\mathbf{a}})$.
\end{algorithmic}
\end{algorithm}

\begin{figure}[t!]
\centerline{\includegraphics[width=70mm]{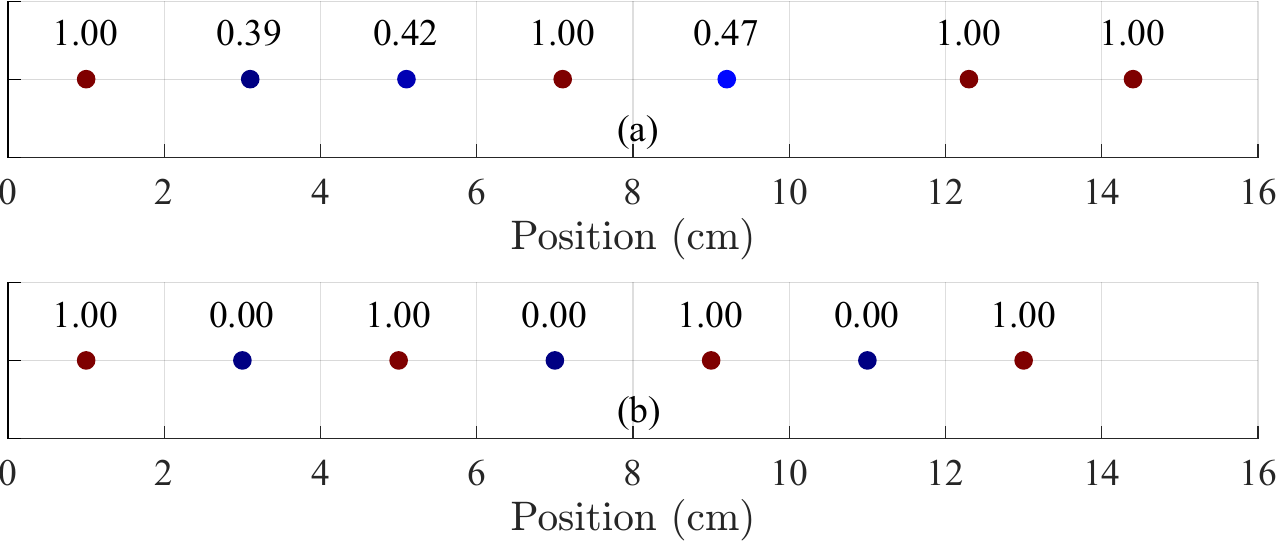}}
	\caption {Illustration of the array geometry and directivity parameters: (a)~LSA-$\textrm{I}$ and (b)~LSA-$\textrm{II}$. The number above each microphone represents its corresponding directivity parameter.}
	\label{exp1}
	\vspace{-16pt}
\end{figure}

\section{Simulations}
In this section, we evaluate and compare the performance of the following two linear superarrays.
\begin{itemize}
\item LSA-\textrm{I}: a superarray optimized using the proposed method, consisting of $M = 7$ microphones, with a minimum microphone spacing of $d_\mathrm{c} = 0.02~\mathrm{m}$ and a maximum aperture of $L = 0.15~\mathrm{m}$.

\item LSA-\textrm{II}: a superarray designed using the method in~\cite{luo2023design}, consisting of 4 omnidirectional microphones and 3 bidirectional microphones, uniformly interspersed with an equal spacing of $0.02~\mathrm{m}$.
\end{itemize}
The simulation frequency range spans from $\omega_\mathrm{L} = 200~\mathrm{Hz}$ to $\omega_\mathrm{H} = 8~\mathrm{kHz}$. The ideal directivity pattern is based on a second-order supercardioid differential beamformer, with parameters set as $\alpha_{2,0} = 0.309$, $\alpha_{2,1} = 0.484$, and $\alpha_{2,2} = 0.207$. The steering direction covers from $ 0^\circ$ to $180^\circ$, and the truncation order is $\mathcal{N} = 2$.

The genetic algorithm parameters are as follows: population size $N_p = 400$, number of generations $N_i = 120$, crossover probability of 0.8, and mutation probability of 0.05.

\begin{figure*}[t!]
\centerline{\includegraphics[width=\textwidth]{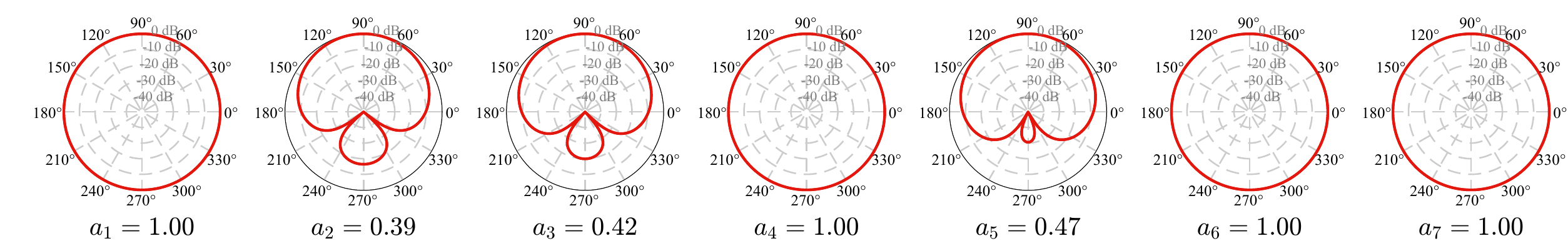}}
	\caption {Directivity patterns of different microphones in LSA-I with the optimized $a_m$ values.}
	\label{exp5}
\end{figure*}

\begin{figure}[t!]
	
\centerline{\includegraphics[width=70mm]{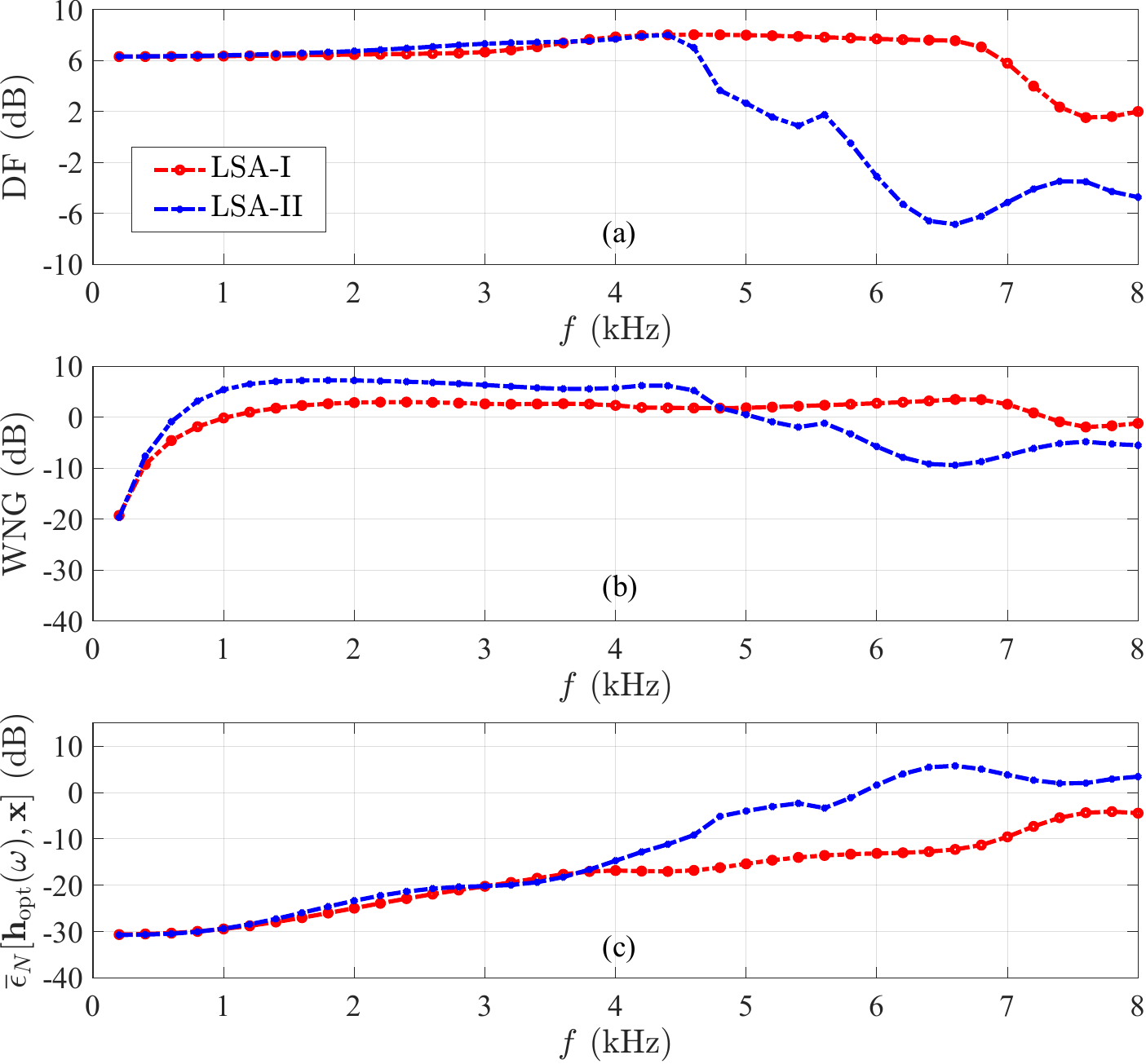}}
	\caption {DF, WNG and beampattern approximation error of LSA-$\textrm{I}$ and LSA-$\textrm{II}$ arrays as functions of $f$. Conditions: $M=7, d_c=0.02~\mathrm{m}, \theta_1=0^\circ, \theta_2=180^\circ$, and $\theta_\mathrm{s}=60^\circ.$}
	\label{exp2}
\end{figure}
\begin{figure}[t!]
	
\centerline{\includegraphics[width=70mm]{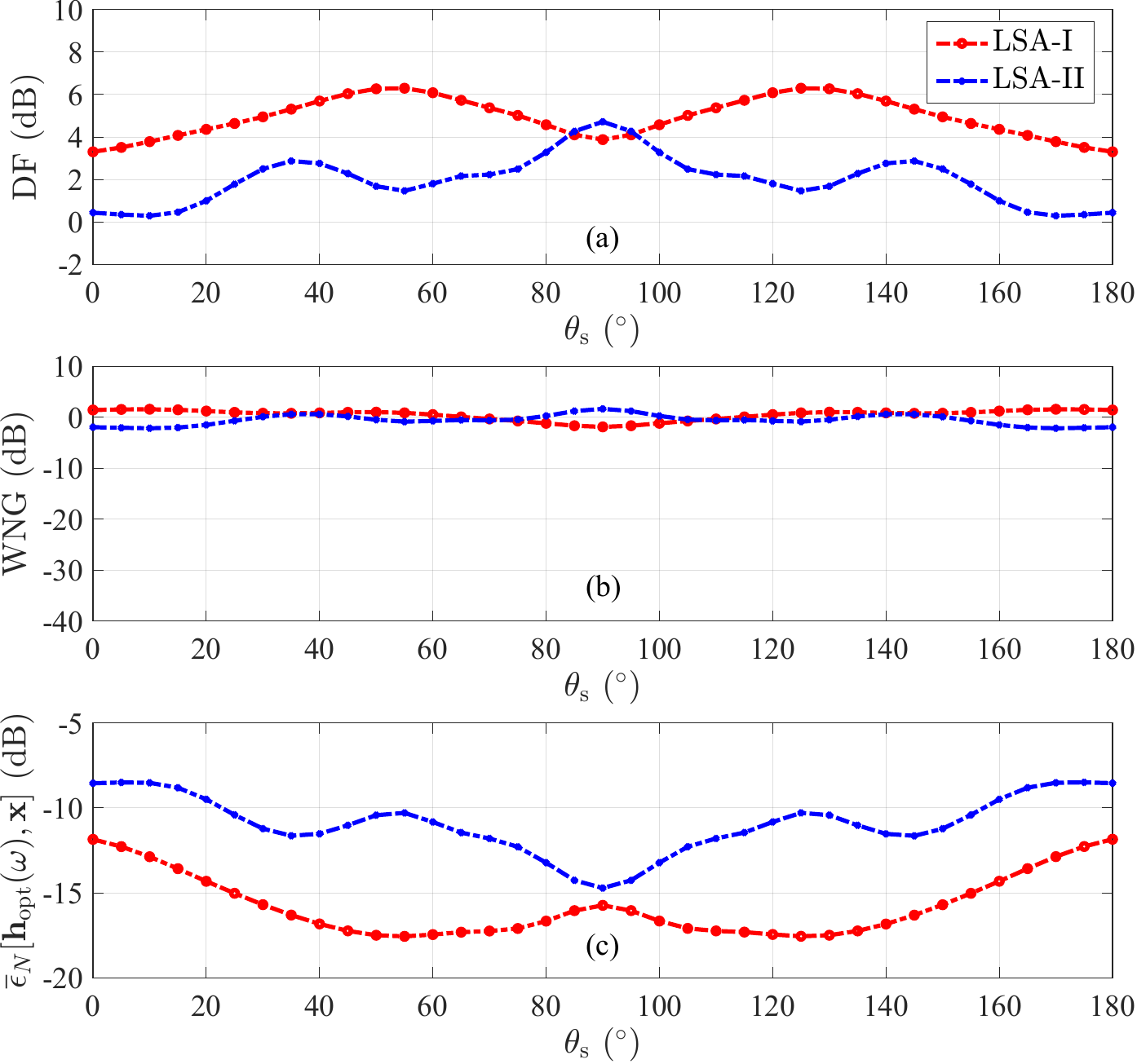}}
	\caption {Average DF, WNG and beampattern approximation error over the full frequency band as functions of $\theta_\mathrm{s}$. Conditions: $M=7, d_c=0.02~\mathrm{m}, \theta_1=0^\circ, \theta_2=180^\circ$, $\omega_\mathrm{L}=200~\mathrm{Hz}$, and $\omega_\mathrm{H}=8000~\mathrm{Hz}$.}
	\label{exp4}
\end{figure}

Figure~\ref{exp1} illustrates the array configurations of (a) LSA-$\textrm{I}$ and (b) LSA-$\textrm{II}$, with the number above each microphone indicating its respective directivity parameters $a_m$. Figure~\ref{exp5} shows the beampatterns for each microphone.

The DF and WNG of LSA-$\textrm{I}$ and LSA-$\textrm{II}$ as functions of frequency $f$ are presented in Fig.~\ref{exp2}, with the steering direction fixed at $\theta_\mathrm{s} = 60^\circ$. It can be observed that while the LSA-$\textrm{II}$ array maintains relatively good DF and WNG performance at low frequencies (below $4$~kHz), its performance deteriorates sharply at higher frequencies. In contrast, the LSA-$\textrm{I}$ array exhibits stable directivity and robustness across the entire frequency band.

Figure~\ref{exp2}(c) shows the approximation error, i.e., a discrete form of~(\ref{opt-2}), which quantifies the deviation between the designed beampatterns of LSA-$\textrm{I}$, LSA-$\textrm{II}$, and the IDP. The results demonstrate that LSA-$\textrm{I}$ consistently achieves a smaller approximation error.

We now compare the performance of the proposed LSA-$\textrm{I}$ in terms of average DF, WNG, and approximation error across the entire frequency band, as functions of the steering direction $\theta_\mathrm{s}$. As shown in Fig.~\ref{exp4}, LSA-$\textrm{I}$ consistently achieves higher DF and lower approximation error than LSA-$\textrm{II}$ for most steering angles, while both arrays demonstrate similar WNG behavior. The notably low average DF of LSA-$\textrm{II}$ stems from its degraded performance at high frequencies, caused by relatively large interelement spacing, as illustrated in Fig.~\ref{exp2}. These results indicate that LSA-$\textrm{I}$ offers improved directivity and consistently smaller approximation errors across various steering directions, thereby providing superior noise suppression.

\begin{figure}[t!]
\centerline{\includegraphics[width=70mm]{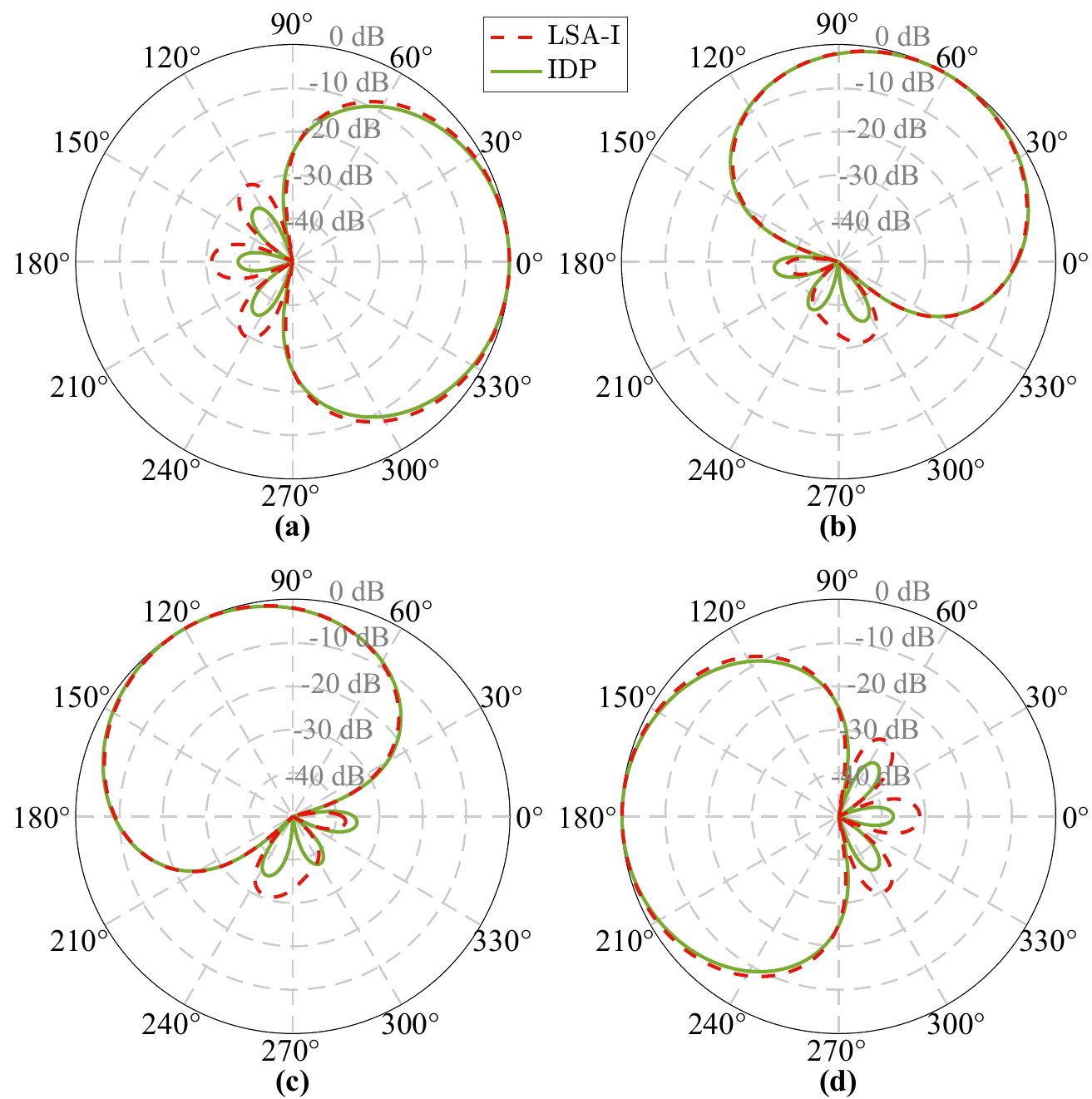}}
	\caption {Beampattern of LSA-$\textrm{I}$ array and the coorsponding IDP at different steering directions $\theta_\mathbf{s}.$}
	\label{exp3}
	\vspace{-20pt}
\end{figure}

Figure~\ref{exp3} shows the beampatterns of the proposed LSA-I at $1$~kHz for various steering directions. The red curves correspond to the LSA-I beampatterns, while the green curves represent the IDP. As the steering direction $\theta_\mathrm{s}$ changes, the overall shape of the beampattern remains consistent, with the mainlobe accurately steered toward the desired direction. These results confirm that the LSA-$\textrm{I}$ designed with the proposed method can flexibly steer across different directions while maintaining consistent spatial responses, demonstrating robust steering capability.

\section{Conclusions}
\label{Sect-Con}
In this paper, we treated microphone array geometry, microphone directivity patterns, and the beamformer as optimizable design variables and presented a generalized framework for the simultaneous optimization of these three parameter categories by minimizing the approximation error between the designed beampattern and an ideal beampattern across the full frequency band and target steering range. A genetic algorithm was developed to minimize the cost function and obtain optimal estimates of the microphone geometry and array element directivity parameters. Simulation results demonstrate that the proposed LSA outperforms existing designs in terms of directivity and beampattern approximation error. These results not only validate the effectiveness of the proposed framework, but also provide valuable insights for future sensor and array development.

 \bibliographystyle{IEEEtran}


\end{document}